\pdfoutput=1

\documentclass[11pt]{article}

\usepackage{amsmath,amsfonts,bm}

\def\eqref#1{equation~\ref{#1}}

\def\1{\bm{1}}

\DeclareMathAlphabet{\mathsfit}{\encodingdefault}{\sfdefault}{m}{sl}
\SetMathAlphabet{\mathsfit}{bold}{\encodingdefault}{\sfdefault}{bx}{n}

\usepackage[final]{acl}

\usepackage{times}
\usepackage{latexsym}

\usepackage[T1]{fontenc}

\usepackage[utf8]{inputenc}

\usepackage{microtype}

\usepackage{inconsolata}

\usepackage{graphicx}

\usepackage{hyperref}       
\usepackage{url}            
\usepackage{booktabs}       
\usepackage{amsfonts}       
\usepackage{nicefrac}       
\usepackage{algorithm}
\usepackage{multirow}
\usepackage{amsmath}
\usepackage{amssymb}
\usepackage{mathtools}
\usepackage{amsthm}
\usepackage{graphicx}
\usepackage{caption}
\usepackage{algorithmic}

\title{Data Poisoning for In-context Learning}

\author{Pengfei He$^{1}$\thanks{Corresponding to hepengf1@msu.edu}, Han Xu$^{2}$, Yue Xing$^1$, Hui Liu$^1$, Makoto Yamada$^3$, Jiliang Tang$^1$\\ 
$^1$Michigan State University  \quad $^2$University of Arizona \\ \quad $^3$ Okinawa Institute of Science and Technology\\ 
}

\begin{document}
\maketitle
\begin{abstract}

In-context learning (ICL) has emerged as a capability of large language models (LLMs), enabling them to adapt to new tasks using provided examples. While ICL has demonstrated its strong effectiveness, there is limited understanding of its vulnerability against potential threats. This paper examines ICL's vulnerability to data poisoning attacks. We introduce \textbf{ICLPoison}, an attacking method specially designed to exploit ICL's unique learning mechanisms by identifying discrete text perturbations that influence LLM hidden states. We propose three representative attack strategies, evaluated across various models and tasks. Our experiments, including those on GPT-4, show that ICL performance can be significantly compromised by these attacks, highlighting the urgent need for improved defense mechanisms to protect LLMs' integrity and reliability.

\end{abstract}





\section{Introduction}
In-context learning (ICL) \citep{brown2020language, min2022rethinking} has emerged as an important capability of large language models (LLMs). Unlike traditional machine learning algorithms that require extensive retraining or fine-tuning to adapt to new tasks \citep{hoi2021online, zhang2021survey, zhuang2020comprehensive}, ICL enables LLMs to make predictions based on a few examples related to a specific task, without changing the model parameters. For example, consider the task of predicting a person's nationality and the prompt consists of examples, e.g. ``Albert Einstein was German; Isaac Newton was English;'', followed by the query ``Thomas Edison was'', an LLM such as GPT-4 will predict ``American'' accurately. The efficiency and flexibility of ICL have drawn significant attention and revolutionized various real-world applications, ranging from LLM-based few-shot healthcare agents \citep{shi2024ehragent} to knowledge tagging in the education domain \citep{li2024knowledge}, where demonstrations are utilized to handle specific tasks.

Despite the success of ICL, studies have shown that the ICL performance is sensitive to certain characteristics of demonstrations, such as the example selection strategy and the quality of examples \citep{wang2023large, min2022rethinking}. Therefore, it naturally raises a question: \textbf{\textit{Is ICL vulnerable to potential data poisoning attacks?}} 
Unlike the traditional data poisoning attack \citep{steinhardt2017certified} where attackers manipulate the training data to corrupt the model trained on the poisoned data, we focus on the perturbations in examples utilized in ICL while keeping the LLMs untouched. 

This paper aims to answer the above question by exploring data poisoning attacks in ICL and uncovering the vulnerability of ICL when facing these attacks.
We consider the standard pipeline of ICL, where examples are randomly selected from a data set for the corresponding downstream task. In terms of the attack, we assume that the adversary deliberately inserts some poisoned examples in this data, and their goal is to ensure that the learning process is adversely affected and the overall ICL prediction performance decreases. This scenario can be both significant and practical. For instance, LLMs and ICL are used in medical systems \citep{shi2024ehragent} for multiple tasks including diagnosis and cost computation. Healthcare providers such as hospitals and physicians may manipulate electronic health records (EHR) to increase revenue \citep{finlayson2018adversarial}. 

Technically, data poisoning in ICL faces both unique challenges specific to ICL and common obstacles in traditional data poisoning. 
First, in contrast to traditional learning algorithms with explicit training objectives, ICL enables LLMs to learn from demonstrations without explicit training \citep{brown2020language, min2022rethinking}. Since traditional poisoning strategies are designed specifically to target the training process and exploit loss functions in conventional models \citep{biggio2012poisoning,steinhardt2017certified,geiping2020witches,he2023sharpness}, 
they are not directly applicable to ICL. Conducting effective data poisoning attacks for ICL requires a thorough understanding of the unique learning mechanism of ICL. Second, similar to traditional attacking methods, data poisoning for ICL also requires creating samples that are imperceptible to humans yet disruptive. These poisoned examples must seamlessly integrate with the other data to harm the learning process subtly. Moreover, one extra challenge of poisoning ICL arises from the discrete vocabulary of language models, making it hard to manipulate inputs for effective disturbance \citep{lei2019discrete, xu2023probabilistic}. 

To tackle the above challenges, we introduce a novel and versatile attacking method, \textbf{ICLPoison}, to exploit the unique learning mechanism of ICL. In particular, previous research \citep{xie2021explanation, hendel2023context, liu2023context, wang2023large} has shown a strong correlation between ICL performance and the hidden states within LLMs. \textbf{ICLPoison} distorts these hidden states through strategic text perturbations, despite the attacker's limited ability to alter only the demonstration examples. We further design three strategies for instantiating and optimizing poisoning attacks under the \textbf{ICLPoison} method. Extensive and comprehensive experiments across various LLMs and tasks demonstrate the effectiveness of our methods, highlighting the vulnerability of ICL. Notably, we have successfully degraded the performance of ICL in advanced models, including GPT-4 (a \textbf{10\%} decrease in ICL accuracy). Our study enhances the understanding of ICL's vulnerability to data poisoning, which helps improve the security and reliability of LLMs.
\section{Related Works}
\subsection{Data Poisoning Attack}
Data poisoning attacks \citep{biggio2012poisoning, steinhardt2017certified} usually occur during the data collection phase of machine learning model training, where the training data is tampered with to induce malicious behaviors in the resulting models \citep{he2023sharpness}. These malicious behaviors include causing a poisoned model to have a poor overall accuracy~\citep{steinhardt2017certified, fowl2021adversarial, huang2021unlearnable}, misclassifying a specified subset of test samples \citep{shafahi2018poison, zhu2019transferable}, or inserting backdoors~\citep{chen2017targeted, gu2019badnets}. 
The data poisoning on traditional NLP models is widely explored. 
CARA \citep{chan2020poison} inserts backdoors into text classifiers by generating poisoned samples by utilizing a conditional adversarially regularized autoencoder; \citep{yang2021careful, schuster2020humpty} poison the embeddings space of NLP models; \citep{chen2021badnl} investigates different triggers to evaluate the effectiveness of backdoor attacks on NLP tasks.

With the development of LLMs, poisoning attacks have also drawn the attention of the whole community, and studies have examined the vulnerability of LLMs against corruption in the pre-training and fine-tuning data. For example, BadPrompt~\citep{cai2022badprompt} inserts backdoors during prompt-tuning and selects effective triggers to maximize the poisoning effect; POISONPROMPT~\citep{yao2024poisonprompt} focuses on a similar setting and leverages a bi-level optimization objective to optimize the trigger while maintaining utility on clean samples; NOTABLE~\citep{mei2023notable} considers the transferability of backdoors and injects backdoors into the encoder to adapt the attack to different downstream tasks and prompting strategies. \citep{wan2023poisoning} further explores the threat during instruction tuning of LLMs and shows that attackers can manipulate model predictions whenever a desired trigger phrase appears in the input. However, these attacks still focus on altering the pre-training or fine-tuning data, and the threat of poisoning examples in ICL remains unexplored. 

\subsection{Attacking In-context Learning (ICL)}

ICL is considered an efficient way to adapt downstream tasks with a few examples, thus the potential safety issue with regard to ICL is of great importance. For instance, \citep{qiang2024hijacking} optimizes adversarial suffixes of examples to mislead the generation of LLMs; \citep{wei2023jailbreak} also employs harmful demonstrations to induce LLMs to produce harmful responses. \citep{kandpal2023backdoor} first explores the threat of backdoor attacks on ICL. In particular, they insert backdoors in pre-training LLMs via fine-tuning on a poisoned dataset. Then when the prompt consists of examples from the target task, the backdoored LLM will misclassify the test sample containing the trigger as the malicious label. Following works \citep{zhao2024universal, zhao2023prompt} propose to directly add triggers into the demonstrations to avoid the computation cost of pre-training or fine-tuning. However, these works require the insertion of obvious triggers that can be easy to detect. In this work, we study more practical scenarios in which imperceptible perturbations are added to demonstrations, and attackers aim to compromise the ICL's overall effectiveness. This presents unique challenges due to the implicit learning mechanism of ICL.
\section{Preliminary}\label{sec:back}

In this section, we introduce ICL and its hidden states, along with the necessary notations.

\textbf{ICL.} 
ICL is a paradigm that allows LLMs to learn tasks given a few examples in the form of demonstrations \citep{brown2020language}. 
To mathematically define the notation of ICL, suppose that we have a pre-trained LLM $f$, which takes in an input prompt $p$ and outputs a response $y$, i.e. $f(p)=y$\footnote{Since the main focus of this paper is not on the generation of LLMs, we adopt the default generation scheme for each LLM.}. Given a task $t\in \mathcal{T}$ from the ICL task set $\mathcal{T}$, we assume $(x,y)\sim \mathcal{D}_t$ where $x\in \mathcal{X}_{t}$ and $y\in \mathcal{Y}_t$. We further assume a prompt set $D_t=\{(x_{i,t}, y_{i,t})\}_{i=1}^N$, where $(x_{i,t}, y_{i,t})\sim \mathcal{D}_t$. 
To conduct an ICL prediction for a query $x^{query}\in \mathcal{X}_{t}$ under task $t$, the user first randomly selects $k$ input-output pairs from $D_t$ and concatenates them as a demonstration $S$, i.e. $S=[(x_{i,t},y_{i,t})]_{i=1}^k$. The demonstration is combined with query $x^{query}$ as an input prompt, and this prompt is sent to the LLM. We define the prediction of ICL as $\hat{y}^{query}_{ICL}=f([S, x^{query}])$. In this work, we consider classification tasks such as sentiment analysis and text classification.

\textbf{Hidden states of ICL.} Extensive studies are conducted to understand the  mechanisms of ICL \citep{xie2021explanation, hendel2023context, von2023transformers, garg2022can, bai2023transformers}. Researchers have demonstrated that the hidden states (represented as $h$), which are defined as the representations of the last token of the input prompt at different layers of the model (as indicated by various studies \citep{hendel2023context, liu2023context, todd2023function}), plays a critical role in ICL. The hidden states can encode the latent concepts in the examples corresponding to the task, which further guides the prediction. In particular, consider a model $f$ composed of $L$ transformer layers, where each layer produces a vector representation $h_l(p,f)\in \mathbb{R}^d$ for the last token of the input prompt $p$ and $l\in[L]$.
It has been observed that the LLMs can make correct predictions conditioning on hidden states $h_l$ extracted from certain intermediate layers \citep{hendel2023context}. 
The hidden states of LLMs provide numerical representations and condense the information from the examples \citep{liu2023context, todd2023function}, which inspires our design of \textbf{ICLPoison}, as illustrated in the next section.

\section{ICLPoison}

In this section, we introduce a novel attacking method, \textbf{ICLPoison}, to conduct data poisoning by distorting hidden states of LLMs during ICL. 


\subsection{Threat Model} \label{sec:threat}

We assume the attacker's goal is to compromise the ICL performance when adapting the downstream tasks using examples from the poisoned dataset as the demonstrations. We assume the attacker can insert poisoned data into the dataset. 
Crucially, attackers are unaware of the details of the ICL process, including the test data, the Large Language Models (LLMs) employed, and specific ICL configurations such as the number of examples and the templates used for demonstrations. Despite these limitations, the attackers can leverage open-source LLMs to generate poisoned data. In real practice, there are many possible scenarios where the attack can insert the poisoned data into the target data set.
For instance, attackers can insert misinformation into the database used in a system where data is collected from various sources \citep{zou2024poisonedrag}; the attackers can also manipulate third-party APIs to insert poisoned examples into the demonstrations \citep{zhao2024attacks}. These threats bring safety and ethical concerns for the real applications of ICL in safety-critic domains like healthcare \citep{shi2024ehragent, joe2021machine}, finance \citep{loukas2023making, paladini2023fraud}, education \citep{li2024knowledge}. 

\subsection{ICLPoison Design} \label{sec:iclpoison}

As highlighted in Section \ref{sec:back}, ICL differs from traditional learning algorithms since it lacks an explicit training objective that can be directly targeted by data poisoning attacks. To address this unique challenge, we draw inspiration from the understanding of hidden states in Section \ref{sec:back} and introduce \textbf{ICLPoison}, a novel data poisoning attack specifically designed for the ICL process. \textbf{ICLPoison} strategically alters examples in demonstrations to distort the hidden states for the poisoning goal.

The details of \textbf{ICLPoison} are as follows. We focus on a surrogate LLM $f$ with $L$ layers and $(x,y)\sim \mathcal{D}_t$. Our objective is to reduce the ICL accuracy for the task $t$ by maximizing the distortion of hidden states of $(x,y)$. We propose perturbing the input $x\sim \mathcal{X}_t$ using a transformation $\delta: \mathcal{X}_t \rightarrow \mathcal{X}_t$ while keeping the label $y$ unchanged. The perturbation $\delta$ must be imperceptible to humans, hence it is constrained within a set $\Delta$ of imperceptible mappings. Details of $\delta$ will be discussed in Section~\ref{sec:algo}.
For each example $(x,y)$ possessed by the attacker, we extract its hidden states. Since the attacker lacks the knowledge of the test data, we use a dummy query $x^{query}_t\sim \mathcal{X}_t$ as a stand-in as suggested in \citep{hendel2023context}. We then concatenate $(x, y)$ with $x^{query}_t$ to create a demonstration and denote $h_{l}(x, f)$ as the representation of the last token in the $l^{th}$ layer of model $f$. Since our focus is on perturbing $x$, we omit $y$ in the following discussion. The representations from all $L$ layers of model $f$ regarding input $x$ are denoted as $H(x,f):=\{h_{l}(x, f)\}_{l=1}^L$, representing the hidden states for $x$ under model $f$. For the perturbed input, the hidden states are $H(\delta_x(x),f)$, with $\delta_x$ being the specific perturbation for $x$.

  To maximize the poisoning effect, we aim to maximize the minimum difference across all layers between the original and perturbed hidden states. To normalize differences across layers with varying scales, we use the normalized $L_2$ norm to measure the distance of the hidden state between the original example and the perturbed one for each layer: $l_d(h_{l}(x,f), h_{l}(\delta_x(x),f))=\|\frac{h_{l}(x,f)}{\|h_{l}(x,f)\|_2}-\frac{h_{l}(\delta_x(x),f)}{\|h_{l}(\delta_x(x),f)\|_2}\|_2$. The distortion between $x$ and $\delta_i(x)$ is further defined as:
\setlength{\abovedisplayskip}{5pt}
\setlength{\belowdisplayskip}{5pt}
\begin{equation}\label{eq:distortion}
\begin{aligned}
    &\mathcal{L}_d(H(x,f), H(\delta_x(x),f))\\
    &=\min\limits_{l\in [L]}l_d(h_{l}(x,f), h_{l}(\delta_x(x),f)).
\end{aligned}
\end{equation}
The attacking objective of \textbf{ICLPoison} is
\begin{equation}\label{eq:iclpoi}
    \max\limits_{\delta_x\in \Delta}\mathcal{L}_d(H(x,f), H(\delta_x(x),f)).
\end{equation}
In other words, $\mathcal{L}_d(H(x,f), H(\delta_x(x),f))$ denotes the minimum changes (or lower bound of the distortion) caused by the perturbation $\delta_x$ across all the layers of the model. This approach ensures that the perturbation $\delta_x$ introduces the most substantial change to the hidden states in the LLM during ICL.
By optimizing the objective in Equation \ref{eq:iclpoi} for each accessible example $(x,y)$, the attacker can create a poisoned example set $D^p_t$.

\subsection{Attacking Algorithms}\label{sec:algo}

To design the perturbation $\delta$, as required in common NLP attacks \citep{ebrahimi2017hotflip, jin2020bert, xu2023probabilistic, li2018textbugger}, $\delta$ is supposed to be imperceptible to humans while effective in manipulating the performance.
In addition to the above requirements, an additional challenge to consider in the optimization is the discrete nature of the objective in Eq.\ref{eq:iclpoi}.
To address these requirements and challenges as well as showcase the versatility of our  method, we introduce three representative perturbations: synonym replacement, character replacement, and adversarial suffix. These methods demonstrate the adaptability of \textbf{ICLPoison} across different levels of text manipulation: Synonym replacement evaluates the word-level vulnerability of ICL and subtly changes the semantics; character replacement involves minimal but precise alterations, 
making it less noticeable to human reviewers (see examples in Appendix \ref{app:example}); and adversarial suffix test token-level vulnerabilities in ICL.
The optimization of these perturbations is managed through a greedy search method, proven effective in similar contexts \citep{lei2019discrete, bao2022towards}.

\textbf{Synonym Replacement.}  This approach involves substituting words in a text with their synonyms, aiming at preserving the semantic meaning and grammatical structure \citep{jin2020bert, xu2023probabilistic}. Within our method, we limit the number of word replacements (denoted as $k$) to maintain the perturbation's imperceptibility. For a text composed of a sequence of $n$ words $x=[w_1,...,w_n]$, $\delta(x)$ is defined as $[s_1,...,s_n]$, where $s_i$ is either a synonym of $w_i$ (if selected for replacement) or remains as $w_i$ (if not replaced). To identify which words are to be replaced and which synonyms are suitable, we follow a strategy similar to \citep{jin2020bert}. We adopt a two-step optimization process. First, we calculate an importance score for each word, selecting the top-$k$ words with the highest scores for replacement. The importance score for word $w_i$ is computed as the distortion (in Eq.\ref{eq:distortion}) before and after deleting $w_i$, expressed as:
\begin{equation}\label{eq:important}
    I_{w_i}= \mathcal{L}_d(H(x,f),H(x_{\text{\textbackslash} w_i},f))
\end{equation}
where $x_{\text{\textbackslash} w_i}$ denotes the text after the removal of $w_i$. In the second step, we use a greedy search method that iteratively replaces each selected word while keeping others fixed \citep{yang2020greedy, lei2019discrete}. For each word, we find synonyms using GloVe embeddings \citep{pennington2014glove}, selecting those with the highest cosine similarity to the original word. Each synonym is temporarily substituted into the text, and the loss function in Eq.\ref{eq:iclpoi} is evaluated. The synonym that maximizes the loss is chosen as the final replacement. This process is repeated for all selected words. We present the detailed algorithm in Algorithm. \ref{algo1}. 

\textbf{Character Replacement.} This method is similar to the synonym replacement approach but focuses on replacing individual characters instead of whole words \citep{ebrahimi2017hotflip, xu2023probabilistic}. When changing only a few letters, this method can be less detectable to humans and maintain the word's pronunciation and basic structures \citep{ebrahimi2017hotflip}.
We limit character replacements to $k$ to keep perturbations subtle. The optimization process involves two steps: first, we calculate an importance score for each character, similar to Eq.\ref{eq:important} but focusing on individual character removal rather than words. The top-$k$ characters with the highest scores are selected for replacement. Second, we use a greedy search strategy, similar to synonym replacement, to find the best replacements for these characters. Note that our character set encompasses uppercase and lowercase letters, digits, punctuation marks, and whitespace, in line with the sets in \citep{kim2016character, ebrahimi2017hotflip}.
The detailed algorithm and its implementation are shown in Algorithm~\ref{algo2} in Appendix ~\ref{app:detail}.

\textbf{Adversarial Suffix.} 
The concept of an adversarial suffix, referred to as adding additional tokens at the end of the original text, has shown considerable effectiveness in misleading LLMs \citep{zou2023universal}. Thus, in addition to synonym and character replacement, we also adapt this perturbation to evaluate the token-level vulnerability of the ICL process. To ensure imperceptible to humans, we restrict the number of additional tokens when adapting to our method. For a given text $x$ that can be tokenized into a sequence of tokens $x=[t_1,...,t_n]$, we define $\delta(x)$ as $[t_1,...,t_n,t'_1,...,t'_k]$ where $t'_1,...,t'_k$ are adversarial suffices. Our goal is to identify the optimal suffixes that maximize the objective in Eq \ref{eq:iclpoi}. We also employ a greedy search approach and iteratively select each suffix token from $t'_1$ to $t'_k$ one by one which results in the maximum increase in the loss. 
The detailed implementation and optimization process for this approach is in Algorithm~\ref{algo3} in Appendix~\ref{app:detail}.
\vspace{-5pt}
\section{Experiments}
We conduct extensive experiments to validate the effectiveness of the proposed method \textbf{ICLPoison}, particularly with three perturbations introduced in Section \ref{sec:algo}. 

\subsection{Experiments setting}

\textbf{Datasets.} We conduct experiments on different types of classification tasks. \textbf{SST2} (2 classes) and \textbf{Cola} (2 classes) are from GLUE dataset \citep{wang2019glue}; \textbf{Emo} \citep{wang2023large} (4 classes) is an emotion classification dataset; \textbf{AG's new} (4 classes) \citep{zhang2015character} is a topic classification dataset; \textbf{Poem} (3 classes) \citep{sheng2020investigating} is a sentiment analysis dataset of poem. Details of these datasets are presented in Appendix \ref{app:dataset}.

\textbf{Models.} We use open-source models including Llama2-7B \citep{touvron2023llama}, Pythia (2.8B, 6.9B) \citep{biderman2023pythia}, Falcon-7B \citep{almazrouei2023falcon}\footnote{Two additional models GPT-J-6B \citep{gpt-j}, MPT-7B \citep{MosaicML2023Introducing} in Appendix \ref{app:adexp}}, and API-only models GPT-3.5 and GPT-4. 

\textbf{Baselines.} Since we are the first to study poisoning attacks in ICL, we compare our methods with clean ICL  and random label flip \citep{min2022rethinking}. For the baseline random label flip, we replace the true label of the example with a random label uniformly selected from the label space.

\textbf{Metrics.} Our main focus is on ICL accuracy. For every dataset, we generate perturbations for examples in the training data and randomly select examples from it to conduct ICL prediction for every sample in the test data. We repeat for 5 runs and report the average ICL accuracy. We also include perplexity scores, which are defined as the average negative log-likelihood of each of the tokens appearing, showing whether the perturbations are imperceptible or not. 

\textbf{Experimental settings.}\footnote{Code can be found in \url{https://anonymous.4open.science/r/ICLPoison-70EE}}. We limit perturbations (also known as budget) to 5 to ensure minimal perceptibility. During the poisoning process, we apply the template: ``\{input\}$\rightarrow$\{output\}\textbackslash{n}\{query\}$\rightarrow$", and extract the hidden states as the representation of the last token ``$\rightarrow$". For evaluation, we adopt the same template and conduct ICL predictions on 5 examples in default. The impact of templates and example numbers is explored in Appendix \ref{app:adexp}. 
\begin{table*}[t]
\tiny
\centering
\captionsetup{font=footnotesize}
\caption{Main results for attacking LLMs. Average ICL accuracy on clean data and poisoned data (5 independent runs) as well as standard error are reported (in percentage), where lower accuracy represents a stronger poisoning effect. The lowest accuracy for each row is highlighted in blue.}
\vspace{-8pt}
\label{tab:1}
\resizebox{0.75\textwidth}{!}
{
\begin{tabular}{c|c|ccccc}
\hline
\hline
\textbf{Model}                           & \textbf{Dataset} & \textbf{Clean}    & \textbf{Random label} & \textbf{Synonym}                         & \textbf{Character}                       & \textbf{Adv suffix}                      \\ \hline
                                         & \textbf{Cola}    & 55.2$\pm$1.8 & 49.3$\pm$2.1     & {\color[HTML]{3531FF} 10.4$\pm$2.1} & 17.6$\pm$1.1                        & 13.8$\pm$1.3                        \\
                                         & \textbf{SST2}    & 82.8$\pm$1.4 & 79.4$\pm$1.9     & {\color[HTML]{3531FF} 19.4$\pm$1.8} & 23.8$\pm$1.9                        & 22.7$\pm$1.6                        \\
                                         & \textbf{Emo}     & 70.3$\pm$2.3 & 39.4$\pm$2.2     & 12.5$\pm$1.1                        & 14.7$\pm$1.4                        & {\color[HTML]{3531FF} 10.4$\pm$1.5} \\
                                         & \textbf{Poem}    & 56.2$\pm$1.8 & 43.1$\pm$2.3     & {\color[HTML]{3531FF} 12.3$\pm$1.5} & 17.9$\pm$1.2                        & 13.8$\pm$1.1                        \\
\multirow{-5}{*}{\textbf{Pythia-6.9B}}   & \textbf{AG}      & 66.5$\pm$2.1 & 47.9$\pm$1.7     & 13.8$\pm$1.3                        & 17.3$\pm$1.5                        & {\color[HTML]{3531FF} 12.9$\pm$1.7} \\ \hline
                                         & \textbf{Cola}    & 63.8$\pm$1.9 & 55.5$\pm$2.0     & 15.3$\pm$1.7                        & 22.7$\pm$2.1                        & {\color[HTML]{3531FF} 13.6$\pm$1.4} \\
                                         & \textbf{SST2}    & 88.6$\pm$1.5 & 82.1$\pm$3.2     & {\color[HTML]{3531FF} 18.5$\pm$2.0} & 26.8$\pm$1.7                        & 20.4$\pm$1.7                        \\
                                         & \textbf{Emo}     & 73.1$\pm$1.3 & 43.6$\pm$1.9     & {\color[HTML]{3531FF} 11.9$\pm$1.8} & 17.5$\pm$1.4                        & 12.7$\pm$1.3                        \\
                                         & \textbf{Poem}    & 62.9$\pm$1.8 & 51.4$\pm$2.3     & 18.1$\pm$1.9                        & 23.3$\pm$1.6                        & {\color[HTML]{3531FF} 17.2$\pm$1.1} \\
\multirow{-5}{*}{\textbf{Llama2-7B}}     & \textbf{AG}      & 73.2$\pm$2.0 & 57$\pm$2.6       & 13.6$\pm$2.2                        & 19.4$\pm$1.3                        & {\color[HTML]{3531FF} 11.9$\pm$1.2} \\ \hline
                                         & \textbf{Cola}    & 65.2$\pm$1.5 & 44.8$\pm$1.7     & 12.7$\pm$1.9                        & 16.5$\pm$1.7                        & {\color[HTML]{3531FF} 10.8$\pm$1.4} \\
                                         & \textbf{SST2}    & 83.8$\pm$2.5 & 83.1$\pm$2.5     & {\color[HTML]{3531FF} 20.1$\pm$1.6} & 25.8$\pm$1.3                        & 22.7$\pm$1.7                        \\
                                         & \textbf{Emo}     & 61.1$\pm$1.7 & 52.6$\pm$1.9     & 10.8$\pm$1.5                        & 14.1$\pm$1.9                        & {\color[HTML]{3531FF} 9.9$\pm$1.1}  \\
                                         & \textbf{Poem}    & 55.2$\pm$1.4 & 42.9$\pm$1.5     & {\color[HTML]{3531FF} 10.5$\pm$1.9} & 17.3$\pm$1.5                        & 13.6$\pm$1.3                        \\
\multirow{-5}{*}{\textbf{Falcon-7B}}     & \textbf{AG}      & 75.2$\pm$1.8 & 50.8$\pm$1.3     & {\color[HTML]{3531FF} 11.2$\pm$2.3} & 14.9$\pm$1.7                        & 12.8$\pm$1.2                        \\ 
\hline\hline
                                         & \textbf{Cola}    & 75.6$\pm$0.7 & 76.3$\pm$0.6     & {\color[HTML]{3531FF} 58.1$\pm$0.5} & 62.6$\pm$0.4                        & 59.7$\pm$0.4                        \\
                                         & \textbf{SST2}    & 93.8$\pm$0.3 & 89.7$\pm$0.5     & 76.8$\pm$0.2                        & 78.3$\pm$0.5                        & {\color[HTML]{3531FF} 74.2$\pm$0.9} \\
                                         & \textbf{Emo}     & 73.8$\pm$0.5 & 72.4$\pm$0.8     & 65.4$\pm$0.4                        & 63.1$\pm$0.7                        & {\color[HTML]{3531FF} 61.3$\pm$0.5} \\
                                         & \textbf{Poem}    & 51.4$\pm$0.9 & 53.3$\pm$0.6     & {\color[HTML]{3531FF} 39.7$\pm$0.6} & 45.2$\pm$0.6                        & 43.9$\pm$0.4                        \\
\multirow{-5}{*}{\textbf{GPT-3.5-turbo}} & \textbf{AG}      & 85.6$\pm$0.3 & 80.7$\pm$0.4     & 76.2$\pm$0.5                        & 73.8$\pm$0.2                        & {\color[HTML]{3531FF} 69.4$\pm$0.7} \\ \hline
                                         & \textbf{Cola}    & 85.8$\pm$0.2 & 82.1$\pm$0.3     & 73.1$\pm$0.5                        & 75.8$\pm$0.3                        & {\color[HTML]{3531FF} 69.6$\pm$0.4} \\
                                         & \textbf{SST2}    & 95.1$\pm$0.4 & 92.5$\pm$0.5     & {\color[HTML]{3531FF} 81.5$\pm$0.2} & 86.1$\pm$0.2                        & 82.3$\pm$0.5                        \\
                                         & \textbf{Emo}     & 84.9$\pm$0.1 & 81.7$\pm$0.2     & 80.9$\pm$0.6                        & {\color[HTML]{3531FF} 78.1$\pm$0.5} & 78.3$\pm$0.4                        \\
                                         & \textbf{Poem}    & 72.4$\pm$0.2 & 63.8$\pm$0.7     & {\color[HTML]{3531FF} 56.7$\pm$0.9} & 60.9$\pm$0.7                        & 57.1$\pm$0.3                        \\
\multirow{-5}{*}{\textbf{GPT-4}}         & \textbf{AG}      & 90.4$\pm$0.3 & 87.3$\pm$0.3     & 83.2$\pm$0.5                        & {\color[HTML]{3531FF} 83.1$\pm$0.4} & 84.7$\pm$0.5                        \\ \hline\hline
\end{tabular}
}
\end{table*}
\vspace{-5pt}
\subsection{Effectiveness of ICLPoison}\label{sec:main}
\begin{figure*}[t]
    \centering
    \includegraphics[width=0.9\textwidth]{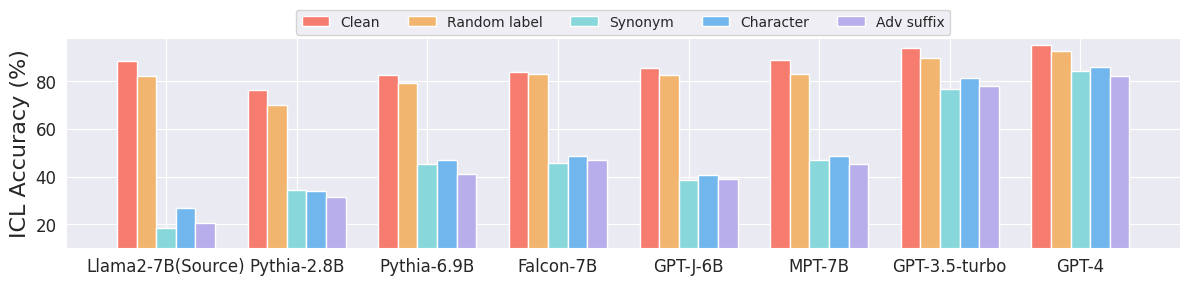}
    \vspace{-8pt}
    \captionsetup{font=footnotesize}
    \caption{Experimental results of transferring poisoned data from Llama2-7B to other models. The Y-axis represents the ICL accuracy (a smaller value represents a stronger poisoning effect), while the X-axis denotes different models.}
    \label{fig:transfer}
    \vspace{-15pt}
\end{figure*}


In this subsection, we first validate the effectiveness of the idea of corrupting the hidden states in \textbf{ICLPoison} across various LLMs and tasks. To eliminate the potential impact on random sampling of poisoned/clean examples, we consider the extreme case of 100\% poisoning rate\footnote{This is an extreme case but still practical. For instance, the victim downloads a whole poisoned data set generated by the attacker for ICL prediction.}.
A more practical attacking scenario with small poisoning rates is discussed in Section~\ref{sec:partial}.

 \textbf{Attacking open-source models.} We first examine open-source models. For each, we craft poisoned samples utilizing the model's own architecture and assess the ICL accuracy. 
Partial results are shown in Table~\ref{tab:1} and full results can be found in Table \ref{tab:open full} in Appendix \ref{app:adexp}. In the table, a lower accuracy indicates a stronger poisoning effect, and the lowest performance is highlighted. One can see that ICL performs well on clean data, especially for the SST2 dataset and Llama2-7B model, achieving more than 88\% accuracy. Besides, the random labeling can only decrease the accuracy marginally, less than 7\%, which is aligned with observations by \citep{min2022rethinking}.
In contrast, our \textbf{ICLPoison}  method significantly reduces ICL accuracy, achieving drops to below 10\% for some models and datasets such as Falcon-7B with the Emo dataset. Notably, the effectiveness of \textbf{ICLPoison} implies that ICL is vulnerable to data poisoning attacks that corrupt the hidden states.

In addition, among the three variants of \textbf{ICLPoison}, synonym replacement and adversarial suffix perturbations pose more severe threats to ICL accuracy compared to character replacement. This disparity may arise because, within the same perturbation budget, different types of perturbations induce varying degrees of change in the hidden states. Character changes typically do not alter the semantic content as significantly as synonym replacements or adversarial suffixes, which can introduce more substantial shifts in the hidden states and disrupt the model's prediction more effectively. We present some poisoned examples from three methods in Appendix \ref{app:example} for human evaluation. 

\noindent\textbf{Attacking API-only models.} For API-only models like GPT-3.5-turbo and GPT-4, we lack direct access to their internal model representations. Therefore, we employ Llama2-7B as a surrogate to generate poisoned samples and assess the ICL accuracy using the provided APIs. The outcomes, detailed in Table \ref{tab:1}, reveal that our approach using Llama2-7B effectively reduces the ICL accuracy of these cutting-edge models by about 10\% for both GPT-3.5 and GPT-4. This not only validates the effectiveness of our method but also confirms its utility in real-world applications with advanced LLMs. Additionally, we observe that compromising such models poses greater challenges than open-source models, potentially due to their larger scale and the use of surrogate models (because of the black-box nature). Furthermore, these models display varying degrees of vulnerability to different perturbation intensities. Notably, GPT-4 exhibits particular susceptibility to character replacement, suggesting a heightened sensitivity to minor textual variations.

\textbf{Transferbility.}
We adopt the Llama2-7B model as the surrogate model and test ICL performance on other models to evaluate the transferability of our poisoning approach across different models, including black-box models. This evaluation validates the transferability of proposed \textbf{ICLPoison} and reveals factors influencing the transferability such as perturbation type and model size. 

Initial results, presented in Figure \ref{fig:transfer}, focus on the SST2 dataset, with a comprehensive analysis of all five datasets available in Appendix \ref{app:adexp}. The results indicate that while the effectiveness of the poisoning attack diminishes when moving from the surrogate (Llama2-7B) to other models, the impact remains significant. Our \textbf{ICLPoison}  method—across its three variants—leads to over a 30\% decrease in accuracy for open-source models. This underscores the substantial threat posed by these attacks.
Our analysis also reveals differences in how various perturbations transfer across models. Synonym replacement and adversarial suffixes demonstrate a stronger poisoning effect compared to character replacement, likely because they introduce more disruption to the model's hidden states. Furthermore, smaller models, such as Pythia-2.8B and GPT-J-6B, are more susceptible to these poisoning examples, whereas larger models exhibit some resistance. This suggests that the effectiveness of our approach can be affected by the size and complexity of the target model.

\begin{figure}[t]
    \centering
    \includegraphics[width=0.47\textwidth]{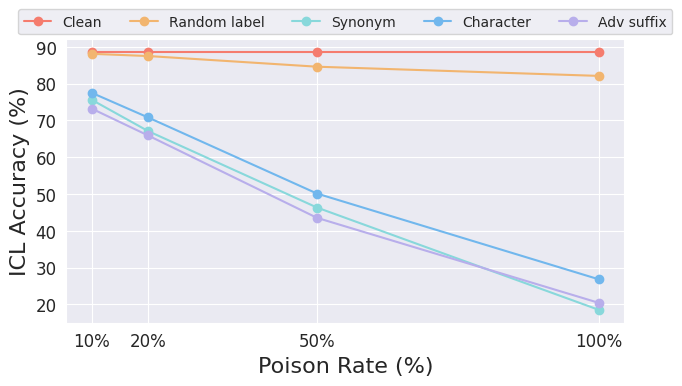}
    \vspace{-8pt}
    \captionsetup{font=footnotesize}
    \caption{Results for practical ICLPoison. The X-axis represents the poison rate while Y-axis represents the ICL accuracy.}
    \label{fig:part}
    \vspace{-18pt}
\end{figure}
The above experimental results confirm that distortions in hidden states can significantly compromise ICL predictions. Synonym replacements and adversarial suffixes cause greater changes in hidden states, leading to stronger poisoning effects. These distortions are also transferable, with smaller models being more affected. These insights underscore the critical link between hidden states and ICL predictions, emphasizing the effectiveness of data poisoning via manipulating hidden states.
\subsection{Practical Attacks}\label{sec:partial}
To align with the threat model in Section \ref{sec:threat}, we consider practical attacking scenarios when attackers only insert some poisoned data into datasets. This usually happens when the victim collects data from various sources and mixes the poisoned data uploaded by the attacker with other clean data \citep{steinhardt2017certified,carlini2024poisoning}. In our experiments, we mix the poisoned data with clean training data, and the poisoning rate varies (10\%, 20\%, 50\%, and 100\%). We test with the Llama2-7B model and GLUE-SST2 dataset. 
The results, shown in Figure \ref{fig:part}, reveal that \textbf{ICLPoison} is still a threat even when only a small set of poisoned data is injected into the clean prompt set. At a small poisoning rate such as 10\% and 20\%, we can observe a significant decrease in ICL performance by over 10\% and 19\% respectively, underscoring the efficacy of our proposed \textbf{ICLPoison}. 
These findings highlight the vulnerability of ICL to subtle data poisoning attacks, where even a limited number of malicious inputs can significantly disrupt the ICL process.
Additionally, similar to the observations in Table \ref{tab:1}, the results show that perturbation strategies like synonym replacement and adversarial suffix, which introduce more pronounced textual changes within a fixed poisoning budget, more severely affect ICL performance compared to character replacement. 

\subsection{Potential Defenses}\label{sec:defense}
Since the attackers are not aware of how victims leverage the poisoned data and some naive defenses may be applied, it is necessary to evaluate the robustness of an attack method. To examine the robustness of \textbf{ICLPoison}, we evaluate three representative defenses: perplexity filtering, paraphrasing \citep{jain2023baseline}, and grammar check. We also take 100\% poisoning rate as the illustration and apply each defense.
\begin{table}[h]
\centering
\captionsetup{font=footnotesize}
\caption{Average perplexity scores for clean and poisoned data with standard error reported. Focus on model Llama2-7B.}
\label{tab:perplexity}
\vspace{-8pt}
\resizebox{0.45\textwidth}{!}
{
\begin{tabular}{c|cccc}
\hline
\textbf{Dataset}         & \textbf{Clean} & \textbf{Synonym} & \textbf{Character} & \textbf{Adv suffix} \\ \hline
\textbf{Cola}       & 4.66$\pm$1.06      & 5.22$\pm$1.00        & 7.37$\pm$0.89          & 9.58$\pm$1.08           \\
\textbf{SST2}       & 5.48$\pm$1.16      & 6.66$\pm$0.73        & 7.45$\pm$0.70          & 7.75$\pm$1.13           \\
\textbf{Emo}             & 5.02$\pm$1.02      & 5.79$\pm$0.67        & 7.27$\pm$0.79          & 7.31$\pm$0.75           \\
\textbf{Poem} & 5.39$\pm$0.85      & 6.32$\pm$0.62        & 8.64$\pm$0.49          & 9.27$\pm$1.09           \\
\textbf{AG}        & 2.37$\pm$0.41      & 3.12$\pm$0.29        & 3.84$\pm$0.50          & 3.68$\pm$0.35           \\ \hline
\end{tabular}
}

\end{table}

\textbf{Perplexity filtering.} The perplexity score measures the average negative log-likelihood of tokens in a text. Perturbations to the original text can increase this score, making them potentially detectable \citep{jain2023baseline}. We present the perplexity scores of generated poisoned data using the Llama2-7B model; higher scores suggest a greater chance of detection and defense. Table \ref{tab:perplexity} displays these scores for poisoned data across various models and datasets, following the calculation method in Section 4.1 of \citep{jain2023baseline}. Among perturbations, synonym replacement results in perplexity scores closest to clean data, while adversarial suffixes yield the highest scores.

\textbf{Paraphrasing.} Paraphrasing is a preprocessing technique that uses a language model to rewrite text, aiming to preserve original meanings while removing adversarial perturbations. We use GPT-4 to paraphrase poisoned data and assess ICL performance on these inputs with the Llama2-7B model. Table \ref{tab:paraphrase} shows that paraphrasing effectively neutralizes adversarial suffixes but largely retains the effects of synonym replacements. To explain this, adversarial suffixes add irrelevant content, while synonym replacements preserve semantic integrity\footnote{Paraphrased text examples are provided in Appendix \ref{app:para}.}.
\begin{table}[h]
\captionsetup{font=footnotesize}
\caption{ICL accuracy when paraphrased by GPT-4. Original results are included in brackets.}
\label{tab:paraphrase}
\vspace{-8pt}
\centering
\resizebox{0.5\textwidth}{!}
{
\begin{tabular}{c|ccccc}
\hline
\textbf{Datasets} & \textbf{Clean} & \textbf{Random label} & \textbf{Synonym}                  & \textbf{Character} & \textbf{Adv suffix} \\ \hline
\textbf{Cola}         & 65.2(63.8)     & 53.9(55.5)            & {\color[HTML]{3531FF} 36.5(15.3)} & 50.6(22.7)         & 58.5(13.6)          \\
\textbf{SST2}         & 83.1(88.6)     & 85.4(82.1)            & {\color[HTML]{3531FF} 52.1(18.5)} & 60.2(26.8)         & 80.2(20.4)          \\
\textbf{Emo}               & 75.5(73.1)     & 48.2(43.6)            & {\color[HTML]{3531FF} 40.7(14.9)} & 48.3(17.5)         & 66.8(12.7)          \\
\textbf{Poem}   & 63.7(62.9)     & 52.1(51.4)            & {\color[HTML]{3531FF} 34.3(18.1)} & 43.7(23.3)         & 55.2(17.2)          \\
\textbf{AG}          & 70.6(73.2)     & 55.7(57)              & {\color[HTML]{3531FF} 38.2(13.6)} & 47.2(19.4)         & 64.3(11.9)          \\ \hline
\end{tabular}
}
\end{table}
\vspace{-10pt}
These findings indicate that defenses enhance ICL's robustness against data poisoning, but their effectiveness varies by perturbation type. Token-level perturbations like adversarial suffixes are easier to counter, whereas word-level 
synonym replacements are more challenging due to subtle semantic changes. Character-level perturbations are moderately detectable but become more severe with larger attack budgets. 
These results highlight the need for more robust ICL defenses.

\textbf{Grammar check defenses.} We evaluate the performance of \textbf{ICLPoison} against typo-corrector \citep{pruthi2019combating} and popular grammar-checking tools--Grammarly and Microsoft Editor. We conduct experiments on SST2 dataset and Llama2-7B model with all three attacking strategies. For typo-corrector, we report the average ICL accuracy; for Grammarly, we report the ratio of correctness errors (CER) over the total number of words; for Microsoft Editor, we report the ratio of spelling errors (SER) and the ratio of grammar errors (GER) over the total number of words. As shown in Table \ref{tab:detection}, the proposed attack can bypass the grammar/typo detection method, especially the synonym replacement strategy. This highlights the stealthiness of \textbf{ICLPoison}. 
\begin{table}[h]
\centering
\captionsetup{font=footnotesize}
\caption{Attacking performance against detection defenses on model Llama2-7B and SST2 dataset.}
\label{tab:detection}
\vspace{-8pt}
\resizebox{0.5\textwidth}{!}
{
\begin{tabular}{c|cccc}
\hline
\textbf{Defenses}                                                                               & \textbf{Clean} & \textbf{Synonym } & \textbf{Adv Suffix} & \textbf{Character} \\ \hline
\begin{tabular}[c]{@{}c@{}}\textbf{Grammarly} \\ \textbf{(CER ($\downarrow$))}\end{tabular}                   & 4.89           & 5.04                         & 11.97                       & 10.31                     \\ \hline
\begin{tabular}[c]{@{}c@{}}\textbf{Microsoft editor} \\ \textbf{(SER/GER ($\downarrow$))}\end{tabular} & 2.72/0.62      & 2.68/0.65                    & 10.08/0.63                  & 7.25/0.68                 \\ \hline
\begin{tabular}[c]{@{}c@{}}\textbf{Typo-corrector} \\ \textbf{(ICL accuracy ($\downarrow$))}\end{tabular}                        & 90.2           & 23.7                         & 41.4                        & 35.8                      \\ \hline
\end{tabular}
}
\end{table}
\vspace{-15pt}
\subsection{Ablation studies}
We present ablation studies with regard to the number of perturbations. Due to the page limit, we provide more ablation studies in Appendix \ref{app:adexp} including token length, prompt template, computation cost, etc. We vary the number of perturbations from 0 to 5, on the SST2 dataset and Llama2-7B model. As shown in Table \ref{tab:num perturbation}, increasing the number of perturbations will enhance the poisoning effect. Moreover, even for a small number of perturbations, such as 2, we can significantly compromise the accuracy of ICL by at least 30\%. 
\begin{table}[h]
\captionsetup{font=footnotesize}
\caption{Ablation on the number of perturbations, model Llama2-7B and SST2 dataset.}
\label{tab:num perturbation}
\vspace{-8pt}
\resizebox{0.5\textwidth}{!}
{
\begin{tabular}{c|cccccc}
\hline
\textbf{Budget}              & \textbf{0(clean)} & \textbf{1} & \textbf{2} & \textbf{3} & \textbf{4} & \textbf{5} \\ \hline
\textbf{Adversarial suffix}  & 88.6              & 69.3       & 50.8       & 34.9       & 25.1       & 20.4       \\
\textbf{Char replacement}    & 88.6              & 73.5       & 57.2       & 46.8       & 32.7       & 26.8       \\
\textbf{Synonym replacement} & 88.6              & 70.1       & 52.9       & 39.1       & 26.4       & 18.5       \\ \hline
\end{tabular}
}
\end{table}
\vspace{-5pt}
\vspace{-10pt}
\section{Conclusion}
In this study, we introduce \textbf{ICLPoison}, a novel method devised to assess the vulnerability of ICL in the face of data poisoning attacks.  We use the dynamics of hidden states in LLMs to craft our attack objectives. Furthermore, we implement our method through three distinct and practical algorithms, each employing a different method of discrete perturbation. Our research exposes previously unidentified susceptibilities of the ICL process to data poisoning. This discovery emphasizes the urgent need for enhancing the robustness of ICL implementations.

\section{Limitations}
In this work, we investigate the vulnerability of in-context learning (ICL) facing data poisoning attacks. In specific, we develop a method \textbf{ICLPoison} to evaluate the vulnerability. The proposed method needs access to the hidden states of LLMs, and thus can not be directly applied to API-only models. We only test on GPT-3.5 and GPT-4, and more API-only models need to be tested. Moreover, attacking methods that only require black-box access are to be investigated to directly attack API-only models. 

\section*{Acknowledgement}
Pengfei He and Jiliang Tang are supported by the National Science Foundation (NSF) under grant numbers CNS2321416, IIS2212032, IIS2212144, IOS2107215, DUE2234015, CNS2246050, DRL2405483 and IOS2035472, the Army Research Office (ARO) under grant number W911NF-21-1-0198, Amazon Faculty Award, JP Morgan Faculty Award, Meta, Microsoft and SNAP.

\bibliography{custom}

\appendix

\appendix
\onecolumn
\section{Appendix}
\subsection{Details of Algorithms} \label{app:detail}
In this section, we present detailed algorithms, including synonym replacement in Algorithm \ref{algo1}, character replacement in Algorithm \ref{algo2} and adversarial suffix in Algorithm \ref{algo3}. 

\noindent\textbf{Algorithm 1}. Algorithm \ref{algo1} describes the whole process of conducting \textbf{ICLPoison} with synonym replacement. For each example $x^p_{i,t}$ in the accessible prompting set $D^p_{t}$, it first selects words to be replaced based on an importance score (step (1)-(3)): the importance score for every word in $x^p_{i,t}$ is computed via Eq.\ref{eq:important} which is the distortion between the text before and after removing the word (step (2)); then the score for every word is sorted in descending order and $k$ words with largest scores are chosen (step (3)). Secondly (step (4)-(8)), we greedily search for the optimal replacement for selected words within their synonyms: first extract $m$ synonyms based on cosine similarity of GloVe embeddings (step (4)); then each selected word is replaced with its synonyms and evaluates the distortion with the original text via Eq. \ref{eq:distortion} (step (5)-(6)); the synonym causing the largest distortion is chosen as the final replacement.

\noindent\textbf{Algorithm 2.} Algorithm \ref{algo2} describes the whole process of conducting \textbf{ICLPoison} with character replacement. For each example $x^p_{i,t}$ in the accessible prompting set $D^p_{t}$, it first selects characters to be replaced based on an importance score (step (1)-(3)): the importance score for every character in $x^p_{i,t}$ is computed via Eq.\ref{eq:important} which is the distortion between the text before and after removing the character (step (2)); then the score for every character is sorted in descending order and $k$ words with largest scores are chosen (step (3)). Secondly (step (4)-(7)), we greedily search for the optimal replacement for selected characters within the character set $C$: each selected word is replaced with characters inside $C$ and evaluates the distortion with the original text via Eq. \ref{eq:distortion} (step (4)-(5)); the character causing the largest distortion is chosen as the final replacement.

\noindent\textbf{Algorithm 3.} Algorithm \ref{algo3} describes the whole process of conducting \textbf{ICLPoison} with adversarial suffix. For each example $x^p_{i,t}$ in the accessible prompting set $D^p_{t}$, it first randomly initializes the $k$ suffices (step (2)). We greedily search for the optimal token for each suffix within the vocabulary set $V$: each suffix is replaced with tokens inside $V$ and evaluates the distortion with the original text via Eq. \ref{eq:distortion} (step (3)-(4)); the character causing the largest distortion is chosen as the final replacement.

\begin{algorithm}[H]
\caption{\textbf{ICLPoison} + Synonym replacement}
\label{algo1}
\textbf{Input} Clean prompting set $D_{t}^p=\{(x^p_{i,t},y^p_{i,t})\}_{i=1}^N$, surrogate model $f$ consisting of $L$ layers, attacking budget $k$, number of synonyms $m$.\\
\textbf{Output} Poisoned prompting set $\{(\delta_i(x^p_{i,t}),y^p_{i,t})\}_{i=1}^N$
\begin{algorithmic}
\FOR {$i=1,...,N$}
\STATE \textbf{Step 1}: \textit{Select words to replace with importance scores}
\STATE (1) Decompose input text $x^p_{i,t}$ into a sequence of words $[w_1,...,w_n]$
\STATE (2) Compute importance score $I_{w_j}$ for each word $w_j$ with Eq. \ref{eq:important}
\STATE (3) Sort scores in descending order $I_{w_{(1)}}\ge I_{w_{(2)}}\ge \cdot \ge I_{w_{(n)}}$ and select top-$k$ words: $w_{(1)},...,w_{(k)}$
\STATE \textbf{Step 2}: \textit{Select optimal synonyms for each selected word}
\FOR {$w \in [w_{(1)},...,w_{(k)}]$}
\STATE (4) Obtain top-$m$ synonyms $[s_{(1)},...,s_{(m)}]$ with highest cosine similarity with $w$ based on GloVe word embeddins.
\FOR {$s\in [s_{(1)},...,s_{(m)}]$}
\STATE (5) Replace $w$ with $s$ obtaining $x'_w=[w_1,...,s,...,w_n]$
\STATE (6) Evaluate the distortion of hidden states after replacement with Eq.\ref{eq:distortion}: $\mathcal{L}_d(H(x^p_{i,t},f), H(x'_w,f))$
\ENDFOR
\STATE (7) Select the synonym causing the largest distortion to replace $w$.
\ENDFOR
\STATE (8) Obtain perturbed input $\delta_i(x^p_{i,t})$
\ENDFOR
\STATE Return poisoned prompting set $\{(\delta_i(x^p_{i,t}),y^p_{i,t})\}_{i=1}^N$
\end{algorithmic}
\end{algorithm}

\begin{algorithm}[H]
\caption{\textbf{ICLPoison} + Character replacement}
\label{algo2}
\textbf{Input} Clean prompting set $D_{t}^p=\{(x^p_{i,t},y^p_{i,t})\}_{i=1}^N$, surrogate model $f$ consisting of $L$ layers, attacking budget $k$, character set $C$.\\
\textbf{Output} Poisoned prompting set $\{(\delta_i(x^p_{i,t}),y^p_{i,t})\}_{i=1}^N$
\begin{algorithmic}
\FOR {$i=1,...,N$}
\STATE \textbf{Step 1}: \textit{Select characters to replace with importance scores}
\STATE (1) Decompose input text $x^p_{i,t}$ into a sequence of characters $[c_1,...,c_n]$
\STATE (2) Compute importance score $I_{c_j}$ for each word $c_j$ with Eq. \ref{eq:important}
\STATE (3) Sort scores in descending order $I_{c_{(1)}}\ge I_{c_{(2)}}\ge \cdot \ge I_{w_{(n)}}$ and select top-$k$ words: $c_{(1)},...,c_{(k)}$
\STATE \textbf{Step 2}: \textit{Select optimal character for each selected character from the whole character set.}
\FOR {$c \in [c_{(1)},...,c_{(k)}]$}
\FOR {$c'\in C$}
\STATE (4) Replace $c$ with $c'$ obtaining $x'_c=[c_1,...,c',...,c_n]$
\STATE (5) Evaluate the distortion of hidden states after replacement with Eq.\ref{eq:distortion}: $\mathcal{L}_d(H(x^p_{i,t},f), H(x'_c,f))$
\ENDFOR
\STATE (6) Select the character causing the largest distortion to replace $c$.
\ENDFOR
\STATE (7) Obtain perturbed input $\delta_i(x^p_{i,t})$
\ENDFOR
\STATE Return poisoned prompting set $\{(\delta_i(x^p_{i,t}),y^p_{i,t})\}_{i=1}^N$
\end{algorithmic}
\end{algorithm}

\begin{algorithm}[H]
\caption{\textbf{ICLPoison} + Adversarial suffix}
\label{algo3}
\textbf{Input} Clean prompting set $D_{t}^p=\{(x^p_{i,t},y^p_{i,t})\}_{i=1}^N$, surrogate model $f$ consisting of $L$ layers, attacking budget $k$, token vocabulary $V$.\\
\textbf{Output} Poisoned prompting set $\{(\delta_i(x^p_{i,t}),y^p_{i,t})\}_{i=1}^N$
\begin{algorithmic}
\FOR {$i=1,...,N$}
\STATE (1) Tokenize text $x^p_{i,t}$ into sequence of tokens $[t_1,...,t_n]$
\STATE (2) Random initialize the adversarial suffix and concatenate with the original text: $\delta(x^p_{i,t})=[t_1,...,t_n,t_1',...,t_k']$
\FOR {$j \in [k]$}
\FOR {$v\in V$}
\STATE (3) Replace $t_j'$ with $v$ obtaining $x'_t=[t_1,...,t_n,t'_1,...,v,...,t'_k]$
\STATE (4) Evaluate the distortion of hidden states after replacement with Eq.\ref{eq:distortion}: $\mathcal{L}_d(H(x^p_{i,t},f), H(x'_t,f))$
\ENDFOR
\STATE (5) Select the token causing the largest distortion to replace $t_j'$.
\ENDFOR
\STATE (6) Obtain perturbed input $\delta_i(x^p_{i,t})$
\ENDFOR
\STATE Return poisoned prompting set $\{(\delta_i(x^p_{i,t}),y^p_{i,t})\}_{i=1}^N$
\end{algorithmic}
\end{algorithm}

\subsection{Details of datasets} \label{app:dataset}
\begin{itemize}
    \item Stanford Sentiment Treebank (\textbf{SST2}) dataset from the GLUE benchmark \citep{wang2019glue} is a sentiment analysis dataset consisting of sentences from movie reviews and human annotations of their sentiment in 2 classes
    \item Corpus of Linguistic Acceptability (\textbf{Cola}) dataset from GLUE is a  linguistic analysis dataset consisting of English acceptability judgments collected from linguistic books, labeled with ``acceptable" or ``unacceptable"
    \item \textbf{Emo} dataset \citep{wang2023large} focuses on emotion classification consisting of Twitter messages labeled in 4 classes
    \item \textbf{AG's new} (AG) corpus \citep{zhang2015character} is a topic classification dataset gathered from news sources and labeled in 4 classes
    \item \textbf{Poem Sentiment} (Poem) \citep{sheng2020investigating} is a sentiment analysis dataset of poem verses from Project Gutenberg, classified into 3 classes
\end{itemize}

\subsection{Additional Experiments} \label{app:adexp}

In this section, we present additional experimental results, including full results on attacking open-source models in Table \ref{tab:open full}, full results of transferability in Table \ref{tab:transfer full}, full results of perplexity scores in Table \ref{tab:perplexity full}, full results on various templates in Table \ref{tab:template app} and the number of examples in Table \ref{tab:num full}.

\noindent\textbf{Attack open-source models}. In Table \ref{tab:open full}, we include more results on additional models such as Pythia-2.8B and MPT-7B. Our observation is consistent with the analysis in Section \ref{sec:main}.

\noindent\textbf{Transferbility}. In Table \ref{tab:transfer full}, results on all 5 datasets are presented, and we notice that the transferability of three perturbations varies. This may be because of the capacity of models and the complexity of datasets. A detailed investigation can be an interesting future direction.

\noindent \textbf{Practical attacks.} We provide the full results of practical attacks with different poison rates in Table \ref{tab:practical}.

\begin{table}[h]
\centering
\caption{Full results of practical attacks.}
\label{tab:practical}
\begin{tabular}{c|ccccc}
\hline
\textbf{Poisoning rate} & \textbf{Clean} & \textbf{Random label} & \textbf{Synonym} & \textbf{Character} & \textbf{Adv suffix} \\ \hline
\textbf{100\%}          & 88.6           & 82.1                  & 18.5             & 26.8               & 20.4                \\
\textbf{50\%}           & 88.6           & 84.6                  & 46.3             & 50.1               & 43.5                \\
\textbf{20\%}           & 88.6           & 87.5                  & 67.1             & 70.8               & 65.9                \\
\textbf{10\%}           & 88.6           & 88.1                  & 75.6             & 77.5               & 73.2                \\ \hline
\end{tabular}
\end{table}

\noindent\textbf{Perplexity scores}. Table \ref{tab:perplexity full} covers perplexity scores on various datasets and models. It is obvious that synonym replacement is more stealthy than the other 2 methods. 

\noindent\textbf{Impact of templates}. We test 3 different templates on various datasets and models. Our results in Table \ref{tab:template app} reveal that our poisoned examples remain effective across templates.

\noindent\textbf{Impact of the number of examples}. Our results about different numbers of examples in Table \ref{tab:num full} show that more examples can improve ICL performance, while also leading to easier manipulation and stronger poisoning effect.
\begin{table}[!ht]
\centering
\caption{Full results on attacking open-source models}
\label{tab:open full}
\resizebox{0.8\textwidth}{!}
{
\begin{tabular}{c|cccccc}
\hline
\textbf{Model}                         & \multicolumn{1}{c|}{\textbf{Dataset}} & \textbf{Clean}    & \textbf{Random label} & \textbf{Synonym}                         & \textbf{Character} & \textbf{Adv suffix}                      \\ \hline
                                       & \multicolumn{1}{c|}{\textbf{Cola}}    & 64.1$\pm$1.6 & 59.4$\pm$1.6     & {\color[HTML]{3531FF} 12.3$\pm$1.2} & 17.6$\pm$1.4  & 14.2$\pm$1.3                        \\
                                       & \multicolumn{1}{c|}{\textbf{SST2}}    & 76.5$\pm$1.5 & 70.2$\pm$1.7     & {\color[HTML]{3531FF} 17.5$\pm$1.1} & 24.3$\pm$1.2  & 18.1$\pm$2.1                        \\
                                       & \multicolumn{1}{c|}{\textbf{Emo}}     & 67.2$\pm$1.3 & 48.1$\pm$2.8     & {\color[HTML]{3531FF} 10.9$\pm$1.8} & 15.7$\pm$1.7  & 12.3$\pm$1.7                        \\
                                       & \multicolumn{1}{c|}{\textbf{Poem}}    & 57.1$\pm$1.7 & 31.8$\pm$1.9     & 10.5$\pm$1.6                        & 16.4$\pm$1.6  & {\color[HTML]{3531FF} 9.7$\pm$1.2}  \\
\multirow{-5}{*}{\textbf{Pythia-2.8B}} & \multicolumn{1}{c|}{\textbf{AG}}      & 59.4$\pm$1.1 & 46.5$\pm$1.7     & 15.7$\pm$1.0                        & 20.3$\pm$1.3  & {\color[HTML]{3531FF} 14.6$\pm$1.4} \\ \hline
                                       & \multicolumn{1}{c|}{\textbf{Cola}}    & 55.2$\pm$1.8 & 49.3$\pm$2.1     & {\color[HTML]{3531FF} 10.4$\pm$2.1} & 17.6$\pm$1.1  & 13.8$\pm$1.3                        \\
                                       & \multicolumn{1}{c|}{\textbf{SST2}}    & 82.8$\pm$1.4 & 79.4$\pm$1.9     & {\color[HTML]{3531FF} 19.4$\pm$1.8} & 23.8$\pm$1.9  & 22.7$\pm$1.6                        \\
                                       & \multicolumn{1}{c|}{\textbf{Emo}}     & 70.3$\pm$2.3 & 39.4$\pm$2.2     & 12.5$\pm$1.1                        & 14.7$\pm$1.4  & {\color[HTML]{3531FF} 10.4$\pm$1.5} \\
                                       & \multicolumn{1}{c|}{\textbf{Poem}}    & 56.2$\pm$1.8 & 43.1$\pm$2.3     & {\color[HTML]{3531FF} 12.3$\pm$1.5} & 17.9$\pm$1.2  & 13.8$\pm$1.1                        \\
\multirow{-5}{*}{\textbf{Pythia-6.9B}} & AG                           & 66.5$\pm$2.1 & 47.9$\pm$1.7     & 13.8$\pm$1.3                        & 17.3$\pm$1.5  & {\color[HTML]{3531FF} 12.9$\pm$1.7} \\ \hline
                                       & \multicolumn{1}{c|}{\textbf{Cola}}    & 63.8$\pm$1.9 & 55.5$\pm$2.0     & 15.3$\pm$1.7                        & 22.7$\pm$2.1  & {\color[HTML]{3531FF} 13.6$\pm$1.4} \\
                                       & \multicolumn{1}{c|}{\textbf{SST2}}    & 88.6$\pm$1.5 & 82.1$\pm$3.2     & {\color[HTML]{3531FF} 18.5$\pm$2.0} & 26.8$\pm$1.7  & 20.4$\pm$1.7                        \\
                                       & \multicolumn{1}{c|}{\textbf{Emo}}     & 73.1$\pm$1.3 & 43.6$\pm$1.9     & {\color[HTML]{3531FF} 11.9$\pm$1.8} & 17.5$\pm$1.4  & 12.7$\pm$1.3                        \\
                                       & \multicolumn{1}{c|}{\textbf{Poem}}    & 62.9$\pm$1.8 & 51.4$\pm$2.3     & 18.1$\pm$1.9                        & 23.3$\pm$1.6  & {\color[HTML]{3531FF} 17.2$\pm$1.1} \\
\multirow{-5}{*}{\textbf{Llama2-7B}}   & \multicolumn{1}{c|}{AG}      & 73.2$\pm$2.0 & 57$\pm$2.6       & 13.6$\pm$2.2                        & 19.4$\pm$1.3  & {\color[HTML]{3531FF} 11.9$\pm$1.2} \\ \hline
                                       & \multicolumn{1}{c|}{\textbf{Cola}}    & 65.2$\pm$1.5 & 44.8$\pm$1.7     & 12.7$\pm$1.9                        & 16.5$\pm$1.7  & {\color[HTML]{3531FF} 10.8$\pm$1.4} \\
                                       & \multicolumn{1}{c|}{\textbf{SST2}}    & 83.8$\pm$2.5 & 83.1$\pm$2.5     & {\color[HTML]{3531FF} 20.1$\pm$1.6} & 25.8$\pm$1.3  & 22.7$\pm$1.7                        \\
                                       & \multicolumn{1}{c|}{\textbf{Emo}}     & 61.1$\pm$1.7 & 52.6$\pm$1.9     & 10.8$\pm$1.5                        & 14.1$\pm$1.9  & {\color[HTML]{3531FF} 9.9$\pm$1.1}  \\
                                       & \multicolumn{1}{c|}{\textbf{Poem}}    & 55.2$\pm$1.4 & 42.9$\pm$1.5     & {\color[HTML]{3531FF} 10.5$\pm$1.9} & 17.3$\pm$1.5  & 13.6$\pm$1.3                        \\
\multirow{-5}{*}{\textbf{Falcon-7B}}   & \multicolumn{1}{c|}{AG}      & 75.2$\pm$1.8 & 50.8$\pm$1.3     & {\color[HTML]{3531FF} 11.2$\pm$2.3} & 14.9$\pm$1.7  & 12.8$\pm$1.2                        \\ \hline
                                       & \multicolumn{1}{c|}{\textbf{Cola}}    & 57.8$\pm$1.3 & 49.1$\pm$2.5     & 13.7$\pm$1.7                        & 17.2$\pm$1.8  & {\color[HTML]{3531FF} 11.8$\pm$0.9} \\
                                       & \multicolumn{1}{c|}{\textbf{SST2}}    & 85.4$\pm$1.6 & 82.8$\pm$2.1     & 14.8$\pm$2.0                        & 18.9$\pm$1.5  & {\color[HTML]{3531FF} 11.4$\pm$1.1} \\
                                       & \multicolumn{1}{c|}{\textbf{Emo}}     & 58.7$\pm$1.1 & 46.2$\pm$1.7     & 11.7$\pm$1.8                        & 13.8$\pm$1.3  & {\color[HTML]{3531FF} 9.6$\pm$0.7}  \\
                                       & \multicolumn{1}{c|}{\textbf{Poem}}    & 57.6$\pm$1.5 & 46.7$\pm$1.4     & 12.6$\pm$2.4                        & 14.2$\pm$2.2  & {\color[HTML]{3531FF} 10.3$\pm$1.3} \\
\multirow{-5}{*}{\textbf{GPT-J-6B}}    & \multicolumn{1}{c|}{\textbf{AG}}      & 63.2$\pm$1.7 & 53.4$\pm$1.9     & {\color[HTML]{3531FF} 11.9$\pm$1.5} & 16.8$\pm$1.8  & 12.5$\pm$1.1                        \\ \hline
                                       & \multicolumn{1}{c|}{\textbf{Cola}}    & 53.4$\pm$1.2 & 45.3$\pm$1.2     & 15.6$\pm$1.6                        & 17.4$\pm$1.9  & {\color[HTML]{3531FF} 14.1$\pm$1.4} \\
                                       & \multicolumn{1}{c|}{\textbf{SST2}}    & 89$\pm$1.5   & 82.9$\pm$2.3     & 20.4$\pm$1.9                        & 25.6$\pm$2.5  & {\color[HTML]{3531FF} 19.8$\pm$1.3} \\
                                       & \multicolumn{1}{c|}{\textbf{Emo}}     & 59.7$\pm$1.3 & 41.8$\pm$1.7     & {\color[HTML]{3531FF} 9.6$\pm$1.5}  & 11.5$\pm$1.6  & 10.4$\pm$0.8                        \\
                                       & \multicolumn{1}{c|}{\textbf{Poem}}    & 69$\pm$1.8   & 56.2$\pm$2.5     & 14.9$\pm$1.4                        & 16.3$\pm$1.5  & {\color[HTML]{3531FF} 12.7$\pm$1.2} \\
\multirow{-5}{*}{\textbf{MPT-7B}}      & \multicolumn{1}{c|}{\textbf{AG}}      & 70.6$\pm$1.6 & 55.3$\pm$1.9     & {\color[HTML]{3531FF} 13.9$\pm$1.7} & 17.1$\pm$1.9  & 15.2$\pm$1.6                        \\ \hline
\end{tabular}
}
\end{table}

\begin{table*}[h!]
\caption{Evaluating data poisoning attacks on different ICL templates. F1, F2, F3 denote 3 different templates, and ICL accuracy on various dataset is reported.}
\label{tab:template app}
\centering
\small
\begin{tabular}{c|c|ccccc}
\hline
\textbf{Dataset}                          & \textbf{ICL format} & \textbf{Clean} & \textbf{Random label} & \textbf{Synonym} & \textbf{Character} & \textbf{Adv suffix} \\ \hline
\multirow{3}{*}{\textbf{Cola}}       & \textbf{F1}                  & 63.8           & 55.5                  & 15.3             & 22.7               & 13.6                \\
                                          & \textbf{F2}                  & 55.1           & 47.5                  & 14.6             & 20.9               & 13.9                \\
                                          & \textbf{F3}                  & 59.5           & 53.3                  & 13.8             & 19.4               & 12.5                \\ \hline
\multirow{3}{*}{\textbf{SST2}}       & \textbf{F1}                  & 88.6           & 82.1                  & 18.5             & 26.8               & 20.4                \\
                                          & \textbf{F2}                  & 92.5           & 84.7                  & 17.9             & 30.6               & 21.7                \\
                                          & \textbf{F3}                  & 90.3           & 79.2                  & 18.2             & 28.5               & 19.3                \\ \hline
\multirow{3}{*}{\textbf{Emo}}             & \textbf{F1}                  & 73.1           & 43.6                  & 14.9             & 17.5               & 12.7                \\
                                          & \textbf{F2}                  & 66.7           & 37.6                  & 11.6             & 12.9               & 9.2                 \\
                                          & \textbf{F3}                  & 69.1           & 39.8                  & 12.8             & 15.6               & 11.3                \\ \hline
\multirow{3}{*}{\textbf{Poem}} & \textbf{F1}                  & 62.9           & 51.4                  & 18.1             & 23.3               & 17.2                \\
                                          & \textbf{F2}                  & 57.1           & 45.2                  & 13.7             & 20.1               & 12.8                \\
                                          & \textbf{F3}                  & 61.9           & 49.5                  & 16.3             & 21.5               & 13.7                \\ \hline
\multirow{3}{*}{\textbf{AG}}        & \textbf{F1}                  & 73.2           & 57.6                  & 13.6             & 19.4               & 11.9                \\
                                          & \textbf{F2}                  & 68.6           & 54.6                  & 11.7             & 18.2               & 10.3                \\
                                          & \textbf{F3}                  & 75.8           & 67.2                  & 14.1             & 19.3               & 12.6                \\ \hline
\end{tabular}
\end{table*}

\begin{table}[]
\centering
\caption{Full results for testing poisoned examples generated by Llama2-7B on other models.}
\label{tab:transfer full}
\resizebox{0.8\textwidth}{!}
{
\begin{tabular}{c|cccccc}
\hline
                                & \textbf{Dataset} & \textbf{Clean} & \textbf{Random label} & \textbf{Synonym}                     & \textbf{Character}                   & \textbf{Adv suffix}                  \\ \hline
                                & \textbf{Cola}    & 64.1  & 59.4         & {\color[HTML]{3531FF} 31.9} & 34.0                        & 34.8                        \\
                                & \textbf{SST2}    & 76.5  & 70.2         & 38.5                        & {\color[HTML]{3531FF} 33.9} & 36.6                        \\
                                & \textbf{Emo}     & 67.2  & 48.1         & {\color[HTML]{3531FF} 26.3} & 32.8                        & 30.3                        \\
                                & \textbf{Poem}    & 57.1  & 31.8         & 33.3                        & 27.1                        & {\color[HTML]{3531FF} 24.3} \\
\multirow{-5}{*}{\textbf{Pythia-2.8B}}   & \textbf{AG}      & 59.4  & 46.5         & 31.7                        & {\color[HTML]{3531FF} 31.6} & 35.6                        \\ \hline
                                & \textbf{Cola}    & 55.2  & 49.3         & {\color[HTML]{3531FF} 26.7} & 34.7                        & 34.2                        \\
                                & \textbf{SST2}    & 82.8  & 79.4         & {\color[HTML]{3531FF} 39.5} & 41.0                        & 41.3                        \\
                                & \textbf{Emo}     & 70.3  & 39.4         & {\color[HTML]{3531FF} 18.9} & 24.1                        & 20.9                        \\
                                & \textbf{Poem}    & 56.2  & 43.1         & 29.0                        & 28.1                        & {\color[HTML]{3531FF} 26.2} \\
\multirow{-5}{*}{\textbf{Pythia-6.9B}}   & \textbf{AG}      & 66.5  & 47.9         & 32.1                        & 33.9                        & {\color[HTML]{3531FF} 27.2} \\ \hline
                                & \textbf{Cola}    & 65.2  & 44.8         & {\color[HTML]{3531FF} 23.4} & 24.8                        & 25.0                        \\
                                & \textbf{SST2}    & 83.8  & 83.1         & 37.7                        & 35.8                        & {\color[HTML]{3531FF} 34.7} \\
                                & \textbf{Emo}     & 61.1  & 52.6         & 28.6                        & 30.5                        & {\color[HTML]{3531FF} 27.1} \\
                                & \textbf{Poem}    & 55.2  & 42.9         & {\color[HTML]{3531FF} 25.1} & 27.3                        & 25.6                        \\
\multirow{-5}{*}{\textbf{Falcon-7B}}     & \textbf{AG}      & 75.2  & 50.8         & 24.7                        & 24.8                        & {\color[HTML]{3531FF} 24.2} \\ \hline
                                & \textbf{Cola}    & 57.8  & 49.1         & 29.7                        & {\color[HTML]{3531FF} 28.5} & 29.1                        \\
                                & \textbf{SST2}    & 85.4  & 82.8         & 31.7                        & 31.8                        & {\color[HTML]{3531FF} 30.2} \\
                                & \textbf{Emo}     & 58.7  & 46.2         & 22.3                        & 24.1                        & {\color[HTML]{3531FF} 19.3} \\
                                & \textbf{Poem}    & 57.6  & 46.7         & 26.7                        & 28.6                        & {\color[HTML]{3531FF} 23.3} \\
\multirow{-5}{*}{\textbf{GPT-J-6B}}      & \textbf{AG}      & 63.2  & 53.4         & {\color[HTML]{3531FF} 29.4} & 29.5                        & 29.8                        \\ \hline
                                & \textbf{Cola}    & 53.4  & 45.3         & 22.6                        & 23.2                        & {\color[HTML]{3531FF} 19.3} \\
                                & \textbf{SST2}    & 89    & 82.9         & 29.8                        & 29.6                        & {\color[HTML]{3531FF} 28.8} \\
                                & \textbf{Emo}     & 59.7  & 41.8         & {\color[HTML]{3531FF} 21.1} & 25.4                        & 25.1                        \\
                                & \textbf{Poem}    & 69    & 56.2         & 26.2                        & 24.5                        & {\color[HTML]{3531FF} 23.5} \\
\multirow{-5}{*}{\textbf{MPT-7B}}        & \textbf{AG}      & 70.6  & 55.3         & 31.0                        & 34.2                        & {\color[HTML]{3531FF} 27.8} \\ \hline
                                & \textbf{Cola}    & 75.6  & 76.3         & {\color[HTML]{3531FF} 58.1} & 62.6                        & 59.7                        \\
                                & \textbf{SST2}    & 93.8  & 89.7         & 76.8                        & 78.3                        & {\color[HTML]{3531FF} 74.2} \\
                                & \textbf{Emo}     & 73.8  & 72.4         & 65.4                        & 63.1                        & {\color[HTML]{3531FF} 61.3} \\
                                & \textbf{Poem}    & 51.4  & 53.3         & {\color[HTML]{3531FF} 39.7} & 45.2                        & 43.9                        \\
\multirow{-5}{*}{\textbf{GPT-3.5-turbo}} & \textbf{AG}      & 85.6  & 80.7         & 76.2                        & 73.8                        & {\color[HTML]{3531FF} 69.4} \\ \hline
                                & \textbf{Cola}    & 85.8  & 82.1         & 73.1                        & 75.8                        & {\color[HTML]{3531FF} 69.6} \\
                                & \textbf{SST2}    & 95.1  & 92.5         & {\color[HTML]{3531FF} 81.5} & 86.1                        & 82.3                        \\
                                & \textbf{Emo}     & 84.9  & 81.7         & 80.9                        & {\color[HTML]{3531FF} 78.1} & 78.3                        \\
                                & \textbf{Poem}    & 72.4  & 63.8         & {\color[HTML]{3531FF} 56.7} & 60.9                        & 57.1                        \\
\multirow{-5}{*}{\textbf{GPT-4}}         & \textbf{AG}      & 90.4  & 87.3         & 83.2                        & {\color[HTML]{3531FF} 83.1} & 84.7                        \\ \hline
\end{tabular}
}
\end{table}

\begin{table}[]
\centering
\caption{Full results for different number of examples in the demonstration. Focus on model Llama2-7B.}
\label{tab:num full}
\resizebox{0.8\textwidth}{!}
{
\begin{tabular}{c|c|ccccc}
\hline
\textbf{Dataset}               & \textbf{num\_examples} & \textbf{Clean} & \textbf{Random label} & \textbf{Synonym} & \textbf{Character} & \textbf{Adv suffix} \\ \hline
\multirow{3}{*}{\textbf{Cola}} & \textbf{3}                      & 63.2           & 59.2                  & 16.5             & 21.8               & 12.8                \\
                               & \textbf{5}                      & 63.8           & 55.5                  & 15.3             & 22.7               & 13.6                \\
                               & \textbf{7}                      & 63.1           & 54.6                  & 14.8             & 22.5               & 13.2                \\ \hline
\multirow{3}{*}{\textbf{SST2}} & \textbf{3}                      & 84.5           & 80.2                  & 14.3             & 29.4               & 19.6                \\
                               & \textbf{5}                      & 88.6           & 82.1                  & 18.5             & 26.8               & 20.4                \\
                               & \textbf{7}                      & 92.7           & 86.1                  & 19.3             & 32.1               & 21.7                \\ \hline
\multirow{3}{*}{\textbf{Emo}}  & \textbf{3}                      & 58             & 35.2                  & 11.5             & 14.6               & 7.9                 \\
                               & \textbf{5}                      & 73.1           & 43.6                  & 14.9             & 17.5               & 12.7                \\
                               & \textbf{7}                      & 79.2           & 47.4                  & 11.7             & 15.8               & 11.3                \\ \hline
\multirow{3}{*}{\textbf{Poem}} & \textbf{3}                      & 61             & 48.6                  & 16.9             & 20.4               & 15.6                \\
                               & \textbf{5}                      & 62.9           & 51.4                  & 18.1             & 23.3               & 17.2                \\
                               & \textbf{7}                      & 66.7           & 55.2                  & 19.5             & 22.6               & 18.3                \\ \hline
\multirow{3}{*}{\textbf{AG}}   & \textbf{3}                      & 66.9           & 52.8                  & 12.9             & 17.5               & 9.5                 \\
                               & \textbf{5}                      & 73.2           & 57                    & 13.6             & 19.4               & 11.9                \\
                               & \textbf{7}                      & 78             & 60.1                  & 13.2             & 18.9               & 11.6                \\ \hline
\end{tabular}
}
\end{table}

\begin{table*}[h!]
\captionsetup{font=footnotesize}
\caption{Perplexity scores for poisoned texts across different models and datasets. A lower value means more logical and fluent expression, and fewer grammar mistakes, thus is more imperceptible to humans.}
\label{tab:perplexity full}
\centering
\resizebox{0.7\textwidth}{!}
{
\begin{tabular}{c|c|cccc}
\hline
                             & \textbf{Dataset} & \textbf{Clean} & \textbf{Synonym} & \textbf{Character} & \textbf{Adv suffix} \\ \hline
\multirow{5}{*}{\textbf{Pythia-2.8B}} & \textbf{Cola}        & 4.87           & 5.15             & 7.38               & 8.35                \\
                             & \textbf{SST2}        & 5.54           & 6.37             & 7.45               & 8.80                 \\
                             & \textbf{Emo}              & 5.46           & 6.07             & 7.27               & 6.81                \\
                             & \textbf{Poem}  & 5.50            & 6.86             & 7.65               & 8.09                \\
                             & \textbf{AG}         & 3.26           & 4.01             & 4.84               & 5.98                \\ \hline
\multirow{5}{*}{\textbf{Pythia-6.9B}} & \textbf{Cola}        & 4.84           & 5.43             & 7.78               & 7.15                \\
                             & \textbf{SST2}        & 5.56          & 5.58             & 7.93               & 7.06                \\
                             & \textbf{Emo}              & 5.41           & 5.92             & 7.49               & 6.45                \\
                             & \textbf{Poem}  & 5.50            & 5.71             & 7.94               & 7.86                \\
                             & \textbf{AG}         & 3.14           & 4.10              & 5.08               & 3.91                \\ \hline
\multirow{5}{*}{\textbf{Llama2-7B}}   & \textbf{Cola}        & 4.66           & 5.22             & 7.37               & 9.58                \\
                             & \textbf{SST2}        & 5.48           & 6.66             & 7.45               & 7.75                \\
                             & \textbf{Emo}              & 5.02           & 5.79             & 7.27               & 7.31                \\
                             & \textbf{Poem}  & 5.39           & 6.32             & 8.64               & 9.27                \\
                             & \textbf{AG}         & 2.37           & 3.12             & 3.84               & 3.68                \\ \hline
\multirow{5}{*}{\textbf{Falcon-7B}}   & \textbf{Cola}        & 5.02           & 5.43             & 7.57               & 7.26                \\
                             & \textbf{SST2}        & 4.76           & 5.33             & 6.84               & 7.12                \\
                             & \textbf{Emo}              & 5.03           & 5.40              & 7.26               & 6.30                 \\
                             & \textbf{Poem}  & 5.53           & 6.02             & 7.80                & 8.07                \\
                             & \textbf{AG}         & 2.67           & 3.49             & 4.50                & 3.64                \\ \hline
\multirow{5}{*}{\textbf{MPT-7B}}      & \textbf{Cola}        & 4.91           & 5.80              & 7.73               & 7.18                \\
                             & \textbf{SST2}        & 5.29           & 5.45             & 6.62               & 6.88                \\
                             & \textbf{Emo}              & 5.12           & 5.47             & 6.40                & 6.22                \\
                             & \textbf{Poem}  & 5.43           & 5.91             & 7.94               & 7.72                \\
                             & \textbf{AG}         & 2.75           & 3.53             & 4.50                & 3.68                \\ \hline
\multirow{5}{*}{\textbf{GPT-J-6B}}    & \textbf{Cola}        & 5.01           & 5.37             & 7.41               & 7.7                 \\
                             & \textbf{SST2}        & 5.06           & 5.35             & 6.92               & 7.81                \\
                             & \textbf{Emo}              & 5.37           & 5.49             & 6.40                & 7.33                \\
                             & \textbf{Poem}  & 5.43           & 5.72             & 8.32               & 8.02                \\
                             & \textbf{AG}         & 3.12           & 4.08             & 4.06               & 5.04                \\ \hline
\end{tabular}
}
\end{table*}

\noindent\textbf{Calibration and selection methods.} We conduct additional experiments with two representative calibration and selection methods, contextual calibration \citep{zhao2021calibrate} and similarity sampling \citep{margatina2023active}, on the SST2 dataset with Llama2-7B. This is because the victims can apply these pre-processing methods. According to the results in Table\ref{tab:cal and sel}, calibration and selection methods can improve the clean accuracy. However, these methods are still vulnerable to the proposed poisoning attack, with a test accuracy of less than 20\%. This may be because the contextual calibration only focuses on the potentially imbalanced distribution of the output label but does not consider the hidden states during calibration. The similarity selection selects demonstrations based on embedding similarity (Sentence-BERT as in \citep{margatina2023active}), which may also be affected by the poisoning.

\begin{table}[h]
\centering
\caption{Test calibration and selection methods on SST2 dataset and Llama2-7B model.}
\label{tab:cal and sel}
\resizebox{0.7\textwidth}{!}
{
\begin{tabular}{c|cc}
\hline
                                & \textbf{Clean} & \textbf{Synonym Attack} \\ \hline
\textbf{Random selection (baseline in our paper)} & 88.6           & 18.5                    \\
\textbf{Contextual calibration}                   & 90.4           & 19.2                    \\
\textbf{Similarity selection}                     & 91.6           & 17.4                    \\ \hline
\end{tabular}
}
\end{table}

\noindent\textbf{Scalability to context length and demonstration size.} We additionally summarize the average token length of each dataset (to represent ‘context length’) used in the experiments. We use the tokenizer of Llama2-7B as the example and present results in Table \ref{tab:context len}. According to the above results, there is no obvious relationship between the attacking performance and the context length. This indicates that our method is robust to the context length.

\begin{table}[h]
\centering
\caption{Average token length of each dataset on Llama2-7B model.}
\label{tab:context len}
\resizebox{0.5\textwidth}{!}
{
\begin{tabular}{c|cc}
\hline
\textbf{Dataset}   & \textbf{Avg token length} & \textbf{attack/clean acc} \\ \hline
\textbf{SST2}      & 14.8                      & 18.5/88.6                 \\
\textbf{COLA}      & 10.4                      & 15.3/63.8                 \\
\textbf{Poem}      & 10.5                      & 18.1/62.9                 \\
\textbf{Emo}       & 18.6                      & 11.9/73.1                 \\
\textbf{AG’s news} & 62.4                      & 13.6/73.2                 \\ \hline
\end{tabular}
}
\end{table}





\noindent\textbf{Computation cost}. We provide the average running time per example (in seconds) when poisoning dataset SST2 on model Llama2-7B with all three methods under different numbers of perturbations (1,3,5 perturbations). As shown in Table \ref{tab:computation}, adversarial suffix and synonym replacement can be time-consuming, while character replacement is much more efficient. This can be due to the size of the search space, as there are more tokens and words than characters. However, since the poisoning process is offline, attackers can take enough time to generate poisoned examples.

\begin{table}[h]
\centering
\caption{Computation cost, SST2, Llama2-7B}
\label{tab:computation}
\resizebox{0.45\textwidth}{!}
{
\begin{tabular}{c|ccc}
\hline
\textbf{Num of perturbation} & \textbf{1} & \textbf{3} & \textbf{5} \\ \hline
\textbf{Adversarial suffix}  & 127        & 369        & 571        \\
\textbf{Char replacement}    & 43         & 118        & 197        \\
\textbf{Synonym replacement} & 71         & 206        & 343        \\ \hline
\end{tabular}
}
\end{table}

\subsection{Poisoned Text Examples} \label{app:example}

In this section, we provide some poisoned examples in Table \ref{tab:example} for human evaluation. The additional tokens (for adversarial suffix) and substitutions (for synonym and character replacement) are highlighted in red. It is obvious that adversarial suffixes can introduce irrelevant or non-sense content to the original text, thus can be easily detected. On the contrast, synonym and character replacements introduce more subtle changes to the text.

\begin{table}[h]
\centering
\caption{Examples of poisoned examples for three type of perturbations.}
\label{tab:example}
\begin{tabular}{|c|p{11cm}|}
\hline
\multirow{5}{*}{\textbf{Adv suffix}} & Singh Hangs Close Despite missing half his fairways and a half-dozen putts inside 12 feet, Vijay Singh grinds out a 1-under 69 to stay within three shots of the lead at the Tour Championship. \textcolor{red}{owing payrollapplyConvert}                                                                               \\ \cline{2-2} 
                                     & ETA Bombs Put Radical Basque Party in Tight Spot The armed Basque separatist group ETA showed it was not a spent force with Friday \#39;s coordinated attacks at five Madrid petrol stations, but it also further isolated the guerrillas \#39; closest political allies, Batasuna. \textcolor{red}{Gould\}\}\$,attempt} \\ \cline{2-2} 
                                     & far from the woods where, when the sun has set,\textcolor{red}{Sup clustering\#absolute}                                                                                                                                                                                                                                 \\ \cline{2-2} 
                                     & Few writers and any playwrights meet in Vienna.\textcolor{red}{ausemployedBlamkAut}                                                                                                                                                                                                                                      \\ \cline{2-2} 
                                     & The tree gradually dropped its fruit to the ground.\textcolor{red}{Anti Africa OH Min==}                                                                                                                                                                                                                                 \\ \hline
\multirow{5}{*}{\textbf{Synonym}}    & Temperatures Dip in Europe, \textcolor{red}{sent(Sending)} Price of Oil Up 4.8 The price of \textcolor{red}{coarse(crude)} oil \textcolor{red}{surging(surged)} 4.8 percent in New York yesterday, the biggest rise in more than five months, on \textcolor{red}{stress(concern)} that cold weather in Europe will draw heating oil from the \textcolor{red}{West(United)} States.           \\ \cline{2-2} 
                                     & \textcolor{red}{should(would)} i \textcolor{red}{queried(ask)} give two \textcolor{red}{qursion(questions)} \textcolor{red}{yeah(yes)} are you \textcolor{red}{marry(married)}                                                                                                                                                                                                               \\ \cline{2-2} 
                                     & \textcolor{red}{Additional(More)} bodies found in Nablus US forces \textcolor{red}{uncovered(discovered)} more bodies under the northern city of Mosul on Friday, \textcolor{red}{evident(apparent)} victims of an \textcolor{red}{bullying(intimidation)} campaign by insurgents against Iraq\textcolor{red}{;(')}s fledgling security forces.                                              \\ \cline{2-2} 
                                     & \textcolor{red}{Vast(Huge)} Black Holes \textcolor{red}{emerged(Formed)} Quickly After Big Bang redOrbit- Incredibly \textcolor{red}{substantial(massive)} black holes had fully \textcolor{red}{developed(matured)} just a billion years after the birth of the \textcolor{red}{cosmos(universe)}, according to two separate studies.                                                       \\ \cline{2-2} 
                                     & can \textcolor{red}{yea(you)} help think \textcolor{red}{search(finding)} girls \textcolor{red}{which(what)} are you \textcolor{red}{state(talking)} about i need \textcolor{red}{rather(first)} girlfriend                                                                                                                                                                                  \\ \hline
\multirow{5}{*}{\textbf{Character}}  & kidney implant hope The first human trial of an artificial 'bio'\textcolor{red}{j(-)}kidney offers \textcolor{red}{c(a)} hop\textcolor{red}{q(e)}\textcolor{red}{Y( )}of a working implant for patients, say ex\textcolor{red}{Q()}perts.                                                                                                                                                    \\ \cline{2-2} 
                                     & oo you are very\textcolor{red}{F( )}lucky haha that \textcolor{red}{s(c)}an be said plu\textcolor{red}{L(s)} i know to maneuver around the roa\textcolor{red}{X(d)}s ' okbuth\textcolor{red}{l(e)}n                                                                                                                                                                                          \\ \cline{2-2} 
                                     & Tennis\textcolor{red}{G(:)} Federer warns rivals Roger Feder\textcolor{red}{g(e)}r believ\textcolor{red}{D(e)}s he is now c\textcolor{red}{f(a)}pable of winning any tournament in tne \textcolor{red}{S(w)}orld.                                                                                                                                                                            \\ \cline{2-2} 
                                     & Zimbabwe curbs rights groups \textcolor{red}{j(Z)}imbabwe's parliam\textcolor{red}{i(e)}nt passes a controversial bill banning international rights gr\textcolor{red}{U(o)}ups from \textcolor{red}{i(w)}or\textcolor{red}{q(k)}ing in the country.                                                                                                                                          \\ \cline{2-2} 
                                     & sentence: t\textcolor{red}{)(h)}e\textcolor{red}{C( )}worthy suc\textcolor{red}{S(c)}essor to a bett\textcolor{red}{n(e)}r to\textcolor{red}{G(m)}orrow                                                                                                                                                                                                                                      \\ \hline
\end{tabular}
\end{table}

\subsection{Paraphrased texts}\label{app:para}
To clarify about paraphrasing, we prompt the model to paraphrase the whole sentences. As a result, it is not a token-level replacement, and it is also hard to quantify how many tokens are changed. We present some examples in Table \ref{tab:paraphrase text}. According to the above table, the paraphrased sentence is different from the original text as it does not only change words but also changes the sentence structure. Besides, as in the third example, when adding some irrelevant content to the input, paraphrasing can remove it. This indicates that the paraphrasing defense is strong compared to attacks.

\begin{table}[h]
\caption{Poisoned texts paraphrased by GPT-4}
\label{tab:paraphrase text}
\begin{tabular}{|p{4.5cm}|p{4.5cm}|p{4.5cm}|}
\hline
\textbf{Original text}& \textbf{Attacked text} &\textbf{Paraphrased text}                                    \\ \hline
Temperatures Dip in Europe, sending Price of Oil Up 4.8 The price of crude oil surged 4.8 percent in New York yesterday, the biggest rise in more than five months, on concern that cold weather in Europe will draw heating oil from the United States.                            
& Temperatures Dip in Europe, sent Price of Oil Up 4.8 The price of coarse oil surging 4.8 percent in New York yesterday, the biggest rise in more than five months, on stress that cold weather in Europe will draw heating oil from the unified states.                                           & European temperatures have fallen, leading to a 4.8\% surging in coarse oil prices in New York yesterday. This marks the largest increase in over five months, driven by stresses that colder conditions in Europe will increase demand for heating oil from the unified states.               \\ \hline
More bodies found in Nablus US forces discovered more bodies under the northern city of Mosul on Friday, apparent victims of an intimidation campaign by insurgents against Iraq’s fledgling security forces.                                    & Additional bodies found in Nablus US forces uncovered more bodies under the northern city of Mosul on Friday, evident victims of an bullying campaign by insurgents against Iraq;s fledgling security forces.                                    & More bodies were discovered in Nablus as US forces unearthed additional victims in the northern city of Mosul on Friday, evidently targeted by insurgents in a campaign of bullying against Iraq's emerging security forces.             \\ \hline
ETA Bombs Put Radical Basque Party in Tight Spot The armed Basque separatist group ETA showed it was not a spent force with Friday \#39;s coordinated attacks at five Madrid petrol stations, but it also further isolated the guerrillas \#39; closest political allies, Batasuna. 
&ETA Bombs Put Radical Basque Party in Tight Spot The armed Basque separatist group ETA showed it was not a spent force with Friday \#39;s coordinated attacks at five Madrid petrol stations, but it also further isolated the guerrillas \#39; closest political allies, Batasuna. Gould\}\}\$,attempt
& The ETA's bombings have placed its key political ally, the radical Basque party Batasuna, in a difficult position. By carrying out synchronized attacks at five Madrid petrol stations on Friday, the armed Basque separatist group demonstrated that it remains a potent force but also increasingly alienated Batasuna from others. 
\\ \hline
\end{tabular}
\end{table}

\end{document}